\documentclass[a4paper]{JHEP3}

\usepackage[english]{babel}
\usepackage{amsmath,amssymb}
\usepackage[dvips]{graphicx}
\usepackage{ifthen,array}
\usepackage[comma,square,sort&compress]{natbib}
\usepackage{amsfonts}
\usepackage{bbm}

\usepackage{dcolumn}% Align table columns on decimal point

\newcommand{\CL}   {C.L.}

\newcommand{\Sol}  {\textsc{sol}}

\newcommand{\Dms}  {\Delta m^2_\Sol}

\newcommand{\Dcq}  {\Delta\chi^2}
\newcommand{\EtAl}  {{\it et al.\/}}
\newcommand{\eps}  {\varepsilon}
\newcommand{\epp}  {\varepsilon'}

\newcommand{\AHEP}{Instituto de F\'{\i}sica Corpuscular --
  C.S.I.C./Universitat de Val{\`e}ncia \\
  Campus de Paterna, Apt 22085,
  E--46071 Val{\`e}ncia, Spain}
\newcommand{\AddrLisb}{%
  Departamento de F\'\i sica and CFTP, Instituto Superior T\'ecnico\\
  Av. Rovisco Pais 1, 1049-001 Lisboa, Portugal }
\newcommand{\Cinvestav}{Departamento de F\'{\i}sica, Centro de
  Investigaci{\'o}n y de Estudios Avanzados del IPN\\ Apdo. Postal
  14-740 07000 Mexico, DF, Mexico}

\title{Are solar neutrino oscillations robust?}

\author{O. G. Miranda \\
\Cinvestav \\
E-mail: \email{Omar.Miranda@fis.cinvestav.mx}}

\author{M. A. T\'ortola \\
\AddrLisb \\
E-mail: \email{mariam@ific.uv.es}}

\author{J. W. F. Valle \\
\AHEP \\
E-mail: \email{valle@ific.uv.es, URL: http://ahep.uv.es/}}

\abstract{ The robustness of the large mixing angle (LMA) oscillation
  (OSC) interpretation of the solar neutrino data is considered in a
  more general framework where non-standard neutrino interactions
  (NSI) are present. Such interactions may be regarded as a generic
  feature of models of neutrino mass.  The 766.3 ton-yr data sample of
  the KamLAND collaboration are included in the analysis, paying
  attention to the background from the reaction $^{13} C(\alpha,
  n)^{16}O$.  Similarly, the latest solar neutrino fluxes from the SNO
  collaboration are included.  In addition to the solution which holds
  in the absence of NSI (LMA-I) there is a ``dark-side'' solution
  (LMA-D) with $\sin^2 \theta_\Sol = 0.70$, essentially degenerate
  with the former, and another light-side solution (LMA-0) allowed
  only at 97\% CL.
  More precise KamLAND reactor measurements will not resolve the
  ambiguity in the determination of the solar neutrino mixing angle
  $\theta_\Sol$, as they are expected to constrain mainly $\Dms$. 
  We comment on the complementary role of atmospheric, laboratory
  (e.~g. CHARM) and future solar neutrino experiments in lifting the
  degeneracy between the LMA-I and LMA-D solutions.
  In particular, we show how the LMA-D solution induced by the
  simplest NSI between neutrinos and down-type-quarks-only is in
  conflict with the combination of current atmospheric data and data
  of the CHARM experiment. We also mention that establishing the issue
  of robustness of the oscillation picture in the most general case
  will require further experiments, such as those involving low energy
  solar neutrinos. }

\keywords{Solar neutrinos; Solar interior; Neutrino interactions;
Neutrino mass and mixing}

\begin{document}

\section{Introduction}
\label{sec:introd}

The very first data of the KamLAND collaboration~\cite{eguchi:2002dm}
have been enough to isolate neutrino oscillations as the correct
mechanism explaining the solar neutrino problem~\cite{pakvasa:2003zv},
indicating also that large mixing angle (LMA) was the right solution.
The 766.3 ton-yr KamLAND data sample strengthens the validity of the
LMA oscillation interpretation of the data~\cite{Araki:2004mb}.

With neutrino experiments now entering the precision
age~\cite{McDonald:2004dd}, the determination of neutrino parameters
and their theoretical impact have become one of the main goals in
astroparticle and high energy physics~\cite{Maltoni:2004ei}.
Now the main efforts should be devoted to the precision determination
of the oscillation parameters and to test for sub-leading
non-oscillation effects such as spin-flavour
conversions~\cite{schechter:1981hw,akhmedov:1988uk} or non-standard
neutrino interactions (NSI, for short)~\cite{Wolfenstein:1977ue}.
 
A quantitative analysis of neutrino oscillations reveals that the
interpretation is relatively robust even taking into account the
possibility of solar density fluctuations in the solar radiative
zone~\cite{Burgess:2003su}, that might arise from magnetic fields
effects~\cite{Burgess:2003fj}, currently unconstrained by
helioseismology.
The robustness of neutrino oscillations in the presence of
spin-flavour conversions induced by non-vanishing neutrino transition
magnetic moments~\cite{Miranda:2003yh} follows from the stringent
limit on anti-neutrinos from the Sun by the KamLAND
collaboration~\cite{Eguchi:2003gg}~\footnote{This does not hold for
the Dirac case, but here the theoretical expectations for magnetic
moments are typically much lower.}.

Here we focus on the case of neutrinos endowed with non-standard
interactions. These are a natural outcome of many neutrino mass
models~\cite{valle:1991pk} and can be of two types: flavour-changing
(FC) and non-universal (NU).

Seesaw-type models leads to a non-trivial structure of the lepton
mixing matrix characterizing the charged and neutral current weak
interactions~\cite{schechter:1980gr}.  This leads to gauge-induced NSI
which may violate lepton flavor and CP even with massless
neutrinos~\cite{mohapatra:1986bd,bernabeu:1987gr,branco:1989bn,rius:1990gk,Deppisch:2004fa}.
Alternatively,  non-standard neutrino interactions may also arise
in models where neutrino masses are ``calculable'' from radiative
corrections~\cite{zee:1980ai,babu:1988ki}.
Finally, in some supersymmetric unified models, the strength of
non-standard neutrino interactions may be a calculable renormalization
effect~\cite{hall:1986dx}.

How sizable are non-standard interactions will be a model-dependent
issue. In some models NSI strengths are too small to be relevant for
neutrino propagation, because they are suppressed by some large scale
and/or restricted by limits on neutrino masses. However, this need not
be the case, and there are interesting models where moderate strength
NSI remain in the limit of light (or even massless)
neutrinos~\cite{mohapatra:1986bd,bernabeu:1987gr,branco:1989bn,rius:1990gk,Deppisch:2004fa}.
Such may occur even in the context of fully unified models like
SO(10)~\cite{Malinsky:2005bi}.

Non--standard interactions may in principle affect neutrino
propagation properties in matter as well as detection cross
sections~\cite{pakvasa:2003zv}. Thus their existence can modify the
solar neutrino signal observed at experiments. They may be
parametrized with the effective low--energy four--fermion operator:
\begin{equation}
    \mathcal{L}_{NSI} = 
    - \epsilon_{\alpha\beta}^{f P} 2\sqrt{2}G_F \left(
    \bar{\nu}_\alpha\gamma_\mu L \nu_\beta
\right)
\left(\bar{f} \gamma^\mu P f \right),
\label{lagrang}
\end{equation}
where  P = L, R and $f$ is a first generation fermion: $e,u,d$. The
coefficients $\eps_{\alpha\beta}^{fP}$ denote the strength of the NSI between 
the neutrinos of flavours $\alpha$ and $\beta$ and the P--handed
component of the fermion $f$.
In the present work, for definiteness, we take for $f$ the down-type
quark.  However, one can also consider the presence of NSI with
electrons and up and down quarks simultaneously. Current limits and
perspectives in the case of NSI with electrons have been reported in the
literature~\cite{NSI-e}.

While strong constraints exist from $\nu_\mu$ interactions with a
down-type quark ($\eps^{d P}_{e\mu} \lesssim 10^{-3}$, $\eps^{d
P}_{\mu\mu} \lesssim 10^{-3}-10^{-2}$) from CHARM and
NuTeV~\cite{Davidson:2003ha}, the constraints for all other NSI
couplings, including those involved in solar neutrino physics, are
rather loose~\cite{Davidson:2003ha,Berezhiani:2001rs}.
Therefore, in our analysis we consider $\eps_{\alpha\mu}^{d P} = 0$
and we concentrate our efforts in the rest of NSI parameters.

For our solar neutrino analysis, we will consider the simplest
approximate two--neutrino picture, which is justified in view of the
stringent limits on $\theta_{13}$~\cite{Maltoni:2004ei} that follow
mainly from reactor neutrino experiments~\cite{apollonio:1999ae}.

The Hamiltonian describing solar neutrino evolution in the presence of
NSI contains, in addition to the standard oscillations term
\begin{equation}
\left(\begin{array}{lc}
-\frac{\Delta m^2}{4E}\cos 2\theta + \sqrt{2}\, G_F N_e ~~ & 
\frac{\Delta m^2}{4E}\sin 2\theta \\
~~~\frac{\Delta m^2}{4E}\sin 2\theta &
\frac{\Delta m^2}{4E}\cos 2\theta
\end{array} \right)
\label{msw-hamiltonian}
\end{equation}
a term $H_\mathrm{NSI}$, accounting for an effective potential induced
by the NSI with matter, which may be written as:
\begin{equation}
    H_\mathrm{NSI} = \sqrt{2} G_F N_d
    \left( \begin{array}{cc}
        0 & \varepsilon \\ \varepsilon & \varepsilon'
    \end{array}\right) \,.
    \label{nsi-hamiltonian}
\end{equation}

Here $\varepsilon$ and $\varepsilon'$ are two effective parameters
that, according to the current bounds discussed above
($\eps_{\alpha\mu}^{f P} \sim 0$), are related with the vectorial
couplings which affect the neutrino propagation by:
\begin{equation}
    \eps = - \sin\theta_{23}\,\eps_{e\tau}^{d V} \qquad
    \epp = \sin^2\theta_{23}\,\eps_{\tau\tau}^{d V} -
    \eps_{ee}^{d V} 
    \label{eff-coup}
\end{equation}
The quantity $N_d$ in Eq.~(\ref{nsi-hamiltonian}) is the number
density of the down-type quark along the neutrino path.
In the more general case, the effective couplings $\eps$ and $\epp$
will contain contributions from the three fundamental fermions and NSI
effects would be important not only in neutrino propagation but also
in the detection process.

It is important to note that the neutrino evolution inside the Sun and
the Earth is sensitive only to the vector component of the NSI,
$\eps_{\alpha\beta}^{d V} = \eps_{\alpha\beta}^{d L} +
\eps_{\alpha\beta}^{d R}$.  The effect of the axial coupling will be
discussed in detail in section 4.

Before introducing our numerical analysis of the solar neutrino data
in the next section, it is worth discussing the analytical formulas
for neutrino survival probability in the constant matter density case,
in order to have a better understanding of the results that will be
shown in the next section.
Recall first that in two--neutrino
oscillations~\cite{deGouvea:2000cq,Gonzalez-Garcia:2000cm} one can,
without loss of generality, restrict the variation of the mixing angle
$\theta$ only to the range $[0,\frac\pi2]$ and still cover the whole
physical space
\footnote{Alternatively one can restrict the angle to the range $0\leq
  \theta \leq \frac \pi4$ if one includes a separate region with
  $\Delta m^2< 0$.  As discussed in \cite{deGouvea:2000cq} it is more
  natural to use $0\leq \theta \leq\frac\pi2$ with a fixed sign of
  $\Delta m^2$. }.
In the adiabatic regime the survival probability can be approximated
by Parke's formula~\cite{Parke:1986jy}
\begin{equation}
    P(\nu_{e} \to \nu_{e}) = \frac{1}{2}\left[ 1 + \cos 2\theta \cos
    2\theta_m \right],    
    \label{prob-parke} 
\end{equation}
where $\theta_m$ is the effective mixing angle at the neutrino
production point inside the sun. In the absence of non-standard
neutrino--matter interactions the mixing angle in matter may be
obtained from the expression
\begin{equation}
    \cos 2\theta_m=
    \frac{\Delta m^2\cos 2\theta - 2\sqrt{2}\,E G_F N_e}
    {\sqrt{\left(\Delta m^2
        \cos 2\theta - 2\sqrt{2}\,E G_F\,N_e \right)^2
        +\left(\Delta m^2\sin 2\theta \right)^2}}\,,
\label{cos-mat}
\end{equation}
In order to explain the deficit of solar neutrinos observed at the
detectors, the neutrino survival probability should satisfy: $P <
0.5$.  According to Eqs.~(\ref{prob-parke}) and (\ref{cos-mat}), this
requirement is not satisfied for $\cos 2\theta < 0$, so that only
vacuum mixing angles in the ``light side'' ($ 0 < \theta < \frac\pi4$)
can give solution to the solar neutrino problem. Indeed this is
confirmed by the results shown in
Fig.~\ref{fig:panels-allowed-region}.

Lets now turn to the case where non-standard interactions are present,
in addition to oscillations.  Within such generalized picture
(OSC+NSI), Eq. (\ref{cos-mat}) is modified to
\begin{equation}
    \cos 2\theta_m=
    \frac{\Delta m^2\cos 2\theta - 2\sqrt{2}\,E G_F (N_e 
      - \varepsilon' N_d)}
    {\left[\Delta m^2\right]_{matter}} ,
    \label{cos-mat-nsi}
\end{equation}
where
\begin{multline}
    \left[
        \Delta m^2
    \right]_{matter}^2 = 
    \left[
        \Delta m^2\cos 2\theta -
        2\sqrt{2}\,E G_F (N_e - \varepsilon' N_d) 
    \right]^2 
    \\
    +
    \left[\Delta m^2
        \sin 2\theta + 4\sqrt{2}\,\varepsilon\, E G_F N_d 
    \right]^2
\end{multline}
Thanks to the presence of the non-universal coupling $\eps'$ one can
obtain $P < 0.5$ even for $\cos 2\theta < 0$ as long as $\eps' >
\frac{2\sqrt{2}\,E G_F N_e + \Delta m^2 |\cos 2\theta|} {2\sqrt{2}\, E
  G_F N_d}$. This makes it possible to explain the solar neutrino data
for values of the vacuum mixing angle in the dark side, for large
enough values of $\varepsilon'$. As we will see below this possibility
will lead to the appearance of the LMA-D solution with $\theta > \frac
\pi4$ and thus to the ambiguous determination of the solar mixing
angle.

\section{The fit}
\label{sec:fit}

\begin{figure}
 \includegraphics[clip,width=0.85\columnwidth,height=6.5cm]{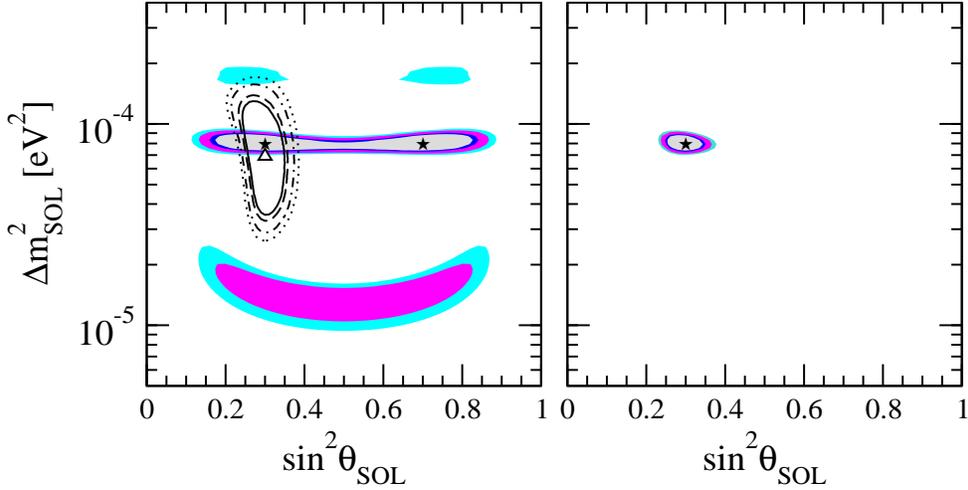}
\caption{\label{fig:panels-allowed-region} 
  90\%, 95\%, 99\% and 99.73\% C.L. allowed regions of the neutrino
  oscillation parameters from the analysis of the latest solar data
  (hollow lines, left panel), and latest KamLAND data (colored
  regions, left panel) and from the combined analysis (right panel).
  }
\end{figure}

Here we reanalyse the robustness of the oscillation interpretation of
the solar neutrino data in the presence of non--standard interactions.
We include the recent SNO data~\cite{Aharmim:2005gt} as well as the
766.3 ton-yr data sample from the KamLAND
collaboration~\cite{Araki:2004mb}, taking into account the background
from the reaction $^{13}C(\alpha,n)^{16}O$.
In order to do this we first calibrate with the results obtained in
the pure oscillation case for the KamLAND--only, solar--only and
combined data samples.
For the KamLAND analysis we use a Poisson statistics as described in
\cite{Fogli:2002au}. Our best fit point is located at $\sin^2
2\theta_\Sol = 0.84$ and $\Dms = 7.9 \times 10^{-5}$
eV~$^2$, in good agreement with the results of
Ref.~\cite{Araki:2004mb}.
For the solar data we include the rates for the Chlorine, Gallex/GNO,
SAGE, as well as the Super-Kamiokande spectrum, SNO day/night spectrum
and SNO salt data, with a total of 84 observables.
We adopt the pull method~\cite{Fogli:2002pt} to fit the data using the
most recent BS05 Standard Solar Model~\cite{Bahcall:2004pz}.
We perform a complete analysis of the solar neutrino data using a
numerical computation for the survival probabilities both in the light
as well as in the dark side of the mixing angle, for values of
$\Dms$ in the range of $10^{-6}$ to $10^{-3}$ eV$^2$ and running
also both $\varepsilon$ and $\varepsilon '$ at the same time in the
range $[-1,1]$.
Our results for $\eps = \eps' = 0$ are shown in
Fig.~\ref{fig:panels-allowed-region}.
The best fit point for this global analysis is given by $\sin^2
\theta_\Sol = 0.29$ and $\Dms = 8.1\times 10^{-5}$
eV~$^2$.
This is in excellent agreement with the results obtained in
\cite{Maltoni:2004ei} for the solar case.
Reassured by this calibration we now turn to the generalized OSC+NSI
picture.
\begin{figure}
\includegraphics[clip,width=0.85\columnwidth, height=6.5cm]{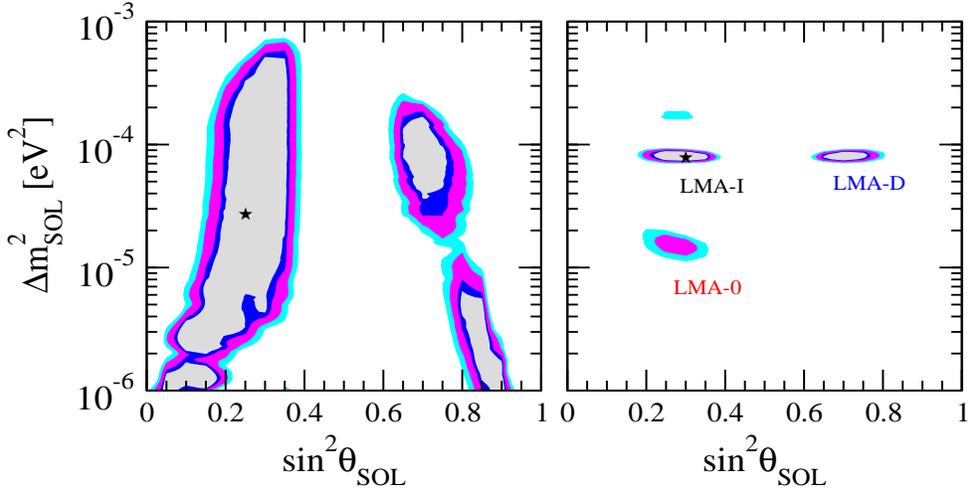}
\caption{\label{fig:panels-allowed-region-nsi} Allowed regions for the
  generalized OSC + NSI case, determined from the latest data: left
  panel corresponds to a solar only analysis, while the right
  panel corresponds to the combined solar+KamLAND analysis.}
\end{figure}
Our results for this case are shown in
Fig.~\ref{fig:panels-allowed-region-nsi} and Table
\ref{tab:bestfit}.
\begin{table}[!t]  
    \begin{center}    
        \begin{tabular}{|c|c|c|c|c|c|}
            \hline      {\rule[-3mm]{0mm}{8mm}  }      & $\sin^2\theta_\Sol$ &
            $\Dms$ [eV$^2$] & $\eps$ & $\eps'$ & $\chi^2$ 
            \\ \hline
            \multicolumn{6}{|c|}{{\rule[-3mm]{0mm}{8mm}       OSC analysis}}
            \\ \hline      LMA-I &  0.29 & 8.1$\times$10$^{-5}$  & -- & -- &
            79.9 
            \\  \hline      
            \multicolumn{6}{|c|}{{\rule[-3mm]{0mm}{8mm} OSC+NSI analysis}}      
            \\ \hline 
            LMA-I & 0.30 & 7.9$\times$10$^{-5}$ & 0 & -0.05 & 79.7 \\ 
            \hline
            LMA-D & 0.70 & 7.9$\times$10$^{-5}$ & -0.15 &
            \hphantom{-}0.90 & 80.2 \\
            \hline 
            LMA-0 & 0.25 & 1.6$\times$10$^{-5}$ & \hphantom{-}0.10 &
            \hphantom{-}0.30 & 86.8 \\  
            \hline    
        \end{tabular}  
    \end{center}  
    \vskip -0.2cm  
    \caption{Best fit solar neutrino oscillation points with and
      without non-standard neutrino interactions.} 
    \label{tab:bestfit}
\end{table}
One sees that, in the light side, we obtain a region of allowed
oscillation parameters larger than in the pure oscillation case, but
more restricted than those obtained in previous OSC+NSI analysis of
Refs.~\cite{Friedland:2004pp,Guzzo:2004ue} due to the effect of the
recent KamLAND data, visible mainly in $\Dms$.
The table gives the parameter best fit values for the OSC and OSC+NSI
fits. For the OSC+NSI analysis the best fit occurs for $\varepsilon =
0.0$ and $\varepsilon' = -0.05$. Clearly the quality of the fit
obtained with and without NSI is comparable, as seen from the $\chi^2$
values given in the last column of the table.
The most remarkable result is, however, the appearance of an
additional solution in the dark side region, which can be
qualitatively understood from the discussion given at the end of
Sect.~\ref{sec:introd}.  This LMA-D solution has $\sin^2 \theta_\Sol =
0.70$ and the same $\Dms$ value as the LMA-I solution and is
significantly better than the LMA-0 OSC+NSI solution of
Ref.~\cite{Friedland:2004pp,Guzzo:2004ue}, as shown in the table. On
the other hand, it is nearly degenerate with the LMA-I solution, as
seen by the $\chi^2$ value.  This solution is characterized by $\eps'=
0.90$, although lower values $\sim$ 0.75 are allowed at 3$\sigma$.
Although embarrassingly large, one sees that such large NSI strength
values are perfectly compatible with all existing solar and reactor
neutrino data, including the small values of the neutrino masses
indicated by current oscillation data. This opens a potentially
physics challenge for upcoming low energy solar neutrino experiments,
such as Borexino. Note that large NSI values could affect also
solar neutrino detection, as considered in~\cite{Berezhiani:2001rt}.
In what follows we give a discussion of the role of other experiments
in probing neutrino properties at the level implied by the above LMA-D
solution.

\section{Constraints on NSI: present and future}
\label{sec:constraints}

As we just saw there are constraints on non-standard neutrino
interaction strength parameters that follow from current solar and
KamLAND data.
The existence of NSI could also potentially affect neutrino-nucleon
scattering and there are laboratory data that potentially constrain
their allowed strength.  Moreover, one must check restrictions that
follow from atmospheric data. Here we discuss their complementarity.

\subsection{Solar and KamLAND}
\label{sec:solar-kamland}

We can derive limits on NSI parameters from solar and KamLAND data by
displaying our $\chi^2$ as a function of the NSI parameters
$\varepsilon$ or $\varepsilon '$ and marginalizing with respect to the
remaining three parameters.
\begin{figure}
\centering
\includegraphics[clip,width=0.85\columnwidth,height=6.5cm]{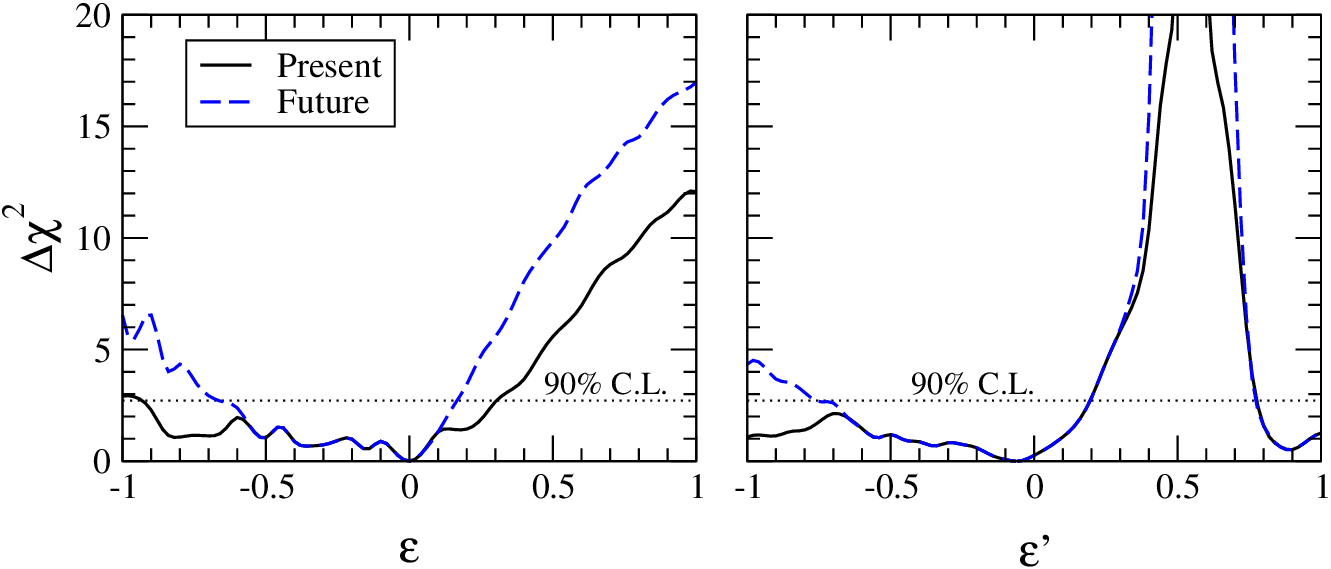}
\caption{\label{fig:epsilon} 
  Constraining NSI parameters: dependence of $\Dcq$ with respect to
  $\varepsilon$ and $\varepsilon'$, illustrating the current
  limits.}
\end{figure}
Figure \ref{fig:epsilon} gives the $\Dcq$ profiles with respect to
$\varepsilon$ and $\varepsilon '$. From here one can determine the
corresponding constraints on $\varepsilon$ and $\varepsilon'$.
We can see that at 90\% C.L. $-0.93 \le \varepsilon \le 0.30$ while
for $\varepsilon '$ the only forbidden region is $[0.20,0.78]$.  Note
that our limits on $\varepsilon '$ are weaker than those of
Ref.~\cite{Guzzo:2004ue}, which apply only to the restricted case
where $\varepsilon=0$. We see that the limits on the strength of
non-standard neutrino interactions are still very poor.
The dashed lines in Fig.~\ref{fig:epsilon} denote the ultimate reach
of this method of constraining NSI parameters (through their effect in
solar neutrino propagation), namely they correspond to the case where
solar neutrino oscillation parameters $\Dms$ and $\theta_\Sol$ are
determined with infinite precision. One sees that in this ideal case
the allowed range narrows down mainly for negative NSI parameter
values.
We conclude that there is substantial room still left for sub--leading
non-standard neutrinos conversions in matter and, moreover, that the
determination of solar neutrino oscillation parameters, especially the
solar mixing angle, is currently ambiguous.  It is unlikely that more
precise reactor measurements by KamLAND will resolve this mixing angle
ambiguity, as they are expected to constrain mainly $\Dms$.

\begin{figure}
\centering
\includegraphics[clip,width=0.8\columnwidth,height=6.5cm]{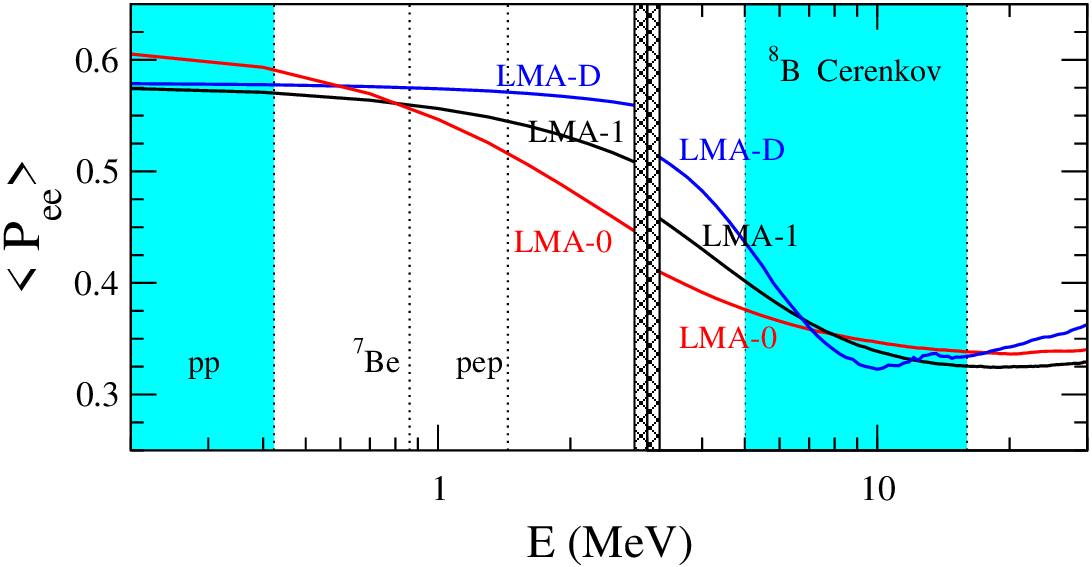}
\caption{\label{fig:spectrum} Predicted neutrino survival probability
for low-energy neutrinos (left) and boron neutrinos (right) at the
best fit points of LMA-I, LMA-D and LMA-0.}
\end{figure}

In Fig.~\ref{fig:spectrum} we present the predicted neutrino survival
probabilities versus energy, from the region of pp neutrinos up to the
high energy solar neutrinos, for the three best--fit points of the
allowed regions found above. One sees that the solutions predict
different rates for the low energy neutrinos (e.g. pp and pep), so 
that future low energy solar neutrino experiments may have a hope of
disentangling these solutions. Similarly, in the region of
boron neutrinos our LMA-D solution also predicts a distortion in the
spectrum that might be detectable at future water Cerenkov experiments
such as UNO or Hyper-K~\cite{UNO}, given the high statistics expected.
With good luck such high statistics experiments may have a window of
opportunity.

\subsection{Laboratory experiments}
\label{sec:labor-exper}

\begin{figure}
    \centering
      \includegraphics[clip,height=6cm,width=0.5\columnwidth]{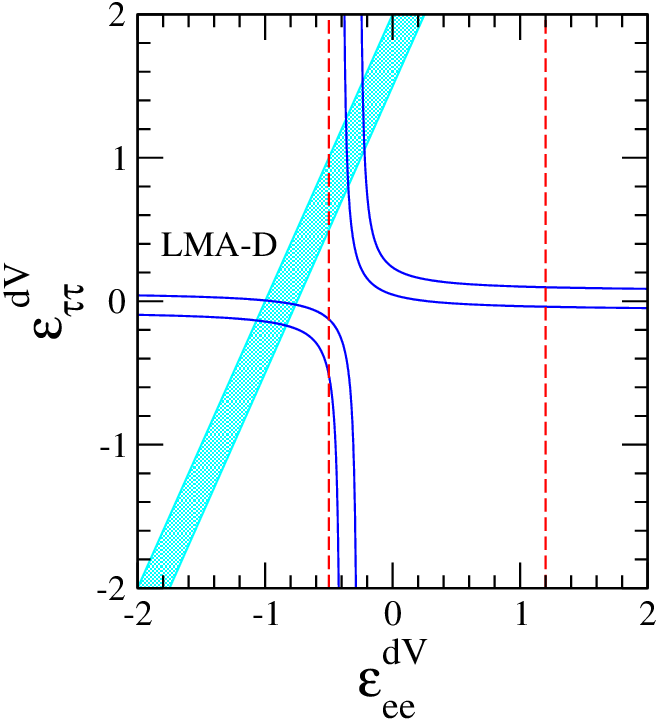}
      \caption{\label{fig:test-atm}  Consistency between
        the $\epp$ coupling required for our LMA-D solution (shaded
        band) and the regions allowed by atmospheric data in the
        analytic approximation of Ref.~\cite{Friedland:2004ah} for
        $\eps_{e\tau}^{d V} = 0.21$ (solid lines). The laboratory  
        constraints are also shown (dashed lines). See the text for a 
        detailed explanation.} 
\end{figure}

The laboratory bounds on the neutrino non-standard interactions with
down-type quarks can be summarized as $|\eps_{\tau e}^{d P} | < 0.5$,
$|\eps_{\tau\tau}^{d R} | < 6$, $|\eps_{\tau\tau}^{d L} | < 1.1$,
$-0.6 < \eps_{ee}^{d R} < 0.5$, $-0.3 < \eps_{ee}^{d L} <
0.3$~\footnote{There is a second branch $0.6 < \eps_{ee}^{d L} < 1.1$
which should be added to the ranges given in
Ref~\cite{Davidson:2003ha}.} , see
e.~g.~Ref~\cite{Davidson:2003ha}. Here we are interested in
vector-like NSI couplings. For the case of $\eps_{ee}^{d V}$, these
bounds can be translated to $-0.5 < \eps_{ee}^{d V} < 1.2$, while for
$\eps_{\tau\tau}^{d V}$ one finds a much wider range. However, we
stress that these bounds have been obtained assuming that only
one parameter is effective at a time. Relaxing this assumption opens
more freedom. This is why we have chosen to indicate them by the
dashed lines in Fig.~\ref{fig:test-atm}.  Assuming maximal mixing in
the 2--3 sector in Eq.~(\ref{eff-coup}), one has
\begin{eqnarray}
  \eps = -0.15  & \to & \eps_{e\tau}^{d V} = 0.21 \\
  \label{eq:test-lab1}    
  \epp =  \hphantom{-}0.90  & \to & \eps_{\tau\tau}^{d V} =
  2~(\eps_{ee}^{dV} + 0.90)
  \label{eq:test-lab2}    
\end{eqnarray}
>From this one can see explicitly that, even taking the above
constraints at face value, they still leave room for our degenerate
dark-side solution with $\epp = 0.90$.

\subsection{Atmospheric data}
\label{sec:atmospheric-data}

Concerning the atmospheric neutrino data, it is known that a large NSI
strengths can originate a suppression of the neutrino oscillation
amplitude. This has indeed been used in a two-neutrino
analysis~\cite{Fornengo:2001pm} in order to obtain relatively strong
bounds on the NSI strength.
However, in a 3--neutrino analysis of atmospheric
data~\cite{Friedland:2004ah} it has been explicitly shown that large
NSI strengths are not excluded. In particular, these authors have
found two specific scenarios where somewhat large NSI strengths can
fit well the experimental data, because their effect will be
indistinguishable from the standard oscillation case, at least at high
and low energies.  Adapting their definitions to our notation, and
using their analytical description~\footnote{This analytical
  comparison holds only at high energies, a more complete check would
  require a numerical analysis to see the effect of intermediate
  energies.}, we obtain the two branches indicated in
Fig.~\ref{fig:test-atm}. One sees that the shaded band corresponding
to our LMA-D solution at 90\% \CL\ (with $\eps_{e\tau}^{dV} = 0.21$)
intersects these branches in two disjoint regions,
suggesting that, indeed, the NSI couplings required by the LMA-D
solution are compatible with the atmospheric neutrino data.
However, in a more complete numerical analysis of atmospheric neutrino
data~\cite{Friedland:2005vy}, it has been shown that values of
$\eps_{\tau\tau}^{dV}$ in the right region are not allowed by
atmospheric data: only the left disjoint region is compatible with 
atmospheric neutrino data. As indicated by the dashed lines in 
Fig.~\ref{fig:test-atm} one can see that $\eps_{ee}^{dV}$ values 
in this region lie outside the range allowed by current laboratory data.
This leads us to conclude that the LMA-D solution induced by the
simplest non-standard interactions of neutrinos with only down-type
quarks is ruled out by its incompatibility with atmospheric and
laboratory data. 
However, one can verify that for the general case where neutrinos have
other NSI couplings one can reconcile the above laboratory
bounds with the parameters required by the LMA-D solution.

\vspace{5mm} Finally, we comment on the magnitude of flavor--changing
NSI of neutrinos. First note that direct constraints on the magnitude
of these couplings do not exist. The only ``bounds'' mentioned in the
literature are obtained from charged lepton flavor violating
processes. While the constraints are very restrictive, they are
theoretically fragile, to the extent that they rely on the assumption
of weak SU(2) symmetry, and can therefore be avoided if one allows for
SU(2) breaking.  Note, however, that the magnitude of the FC neutrino
NSI required for our dark--side solution is quite small. Futuristic
proposals for improving these constraints with coherent neutrino
scattering off nuclei have been already
discussed~\cite{Barranco:2005yy}.

\section{A comment on axial NSI couplings}
\label{sec:axial}

Before we conclude, let us mention that, up to now we have only
considered the effects of the NSI on the neutrino propagation through
the Earth and solar interior. These effects appear as a result of the
vectorial couplings of neutrinos with down-type quarks. An axial
component of the NSI coupling could give rise to a non--standard
contribution to the NC cross section detection at the SNO experiment.
As already noted in Ref.\cite{Davidson:2003ha}, the SNO-NC signal will
be modified as:
\begin{equation}
\phi_{NC} \sim f_B (1 + 2\eps_A) 
\end{equation}
where
\begin{equation}
\eps_A = -\sum_{\alpha = e,\mu,\tau}\left< P_{e\alpha} \right> _{NC} \eps_{\alpha\alpha}^{d A}
\end{equation}
with $\eps_{\alpha\alpha}^{d A} = \eps_{\alpha\alpha}^{d L} -
\eps_{\alpha\alpha}^{d R}$ being the couplings which enter into the
effective Lagrangian. Thus, $\eps_A$ is independent of the effective
couplings $\eps$ and $\epp$ defined in Eq.~(\ref{eff-coup}).  In the
analysis performed so far we have assumed $\eps_A = 0$. This
assumption is well justified due to the good agreement between the SNO
NC measurement and the SSM prediction for the boron flux. 
However, we now relax this assumption and include the effect of 
the new parameter $\eps_A$ in our analysis. 

\begin{figure}
\centering
\includegraphics[clip,width=0.85\columnwidth,height=6cm]{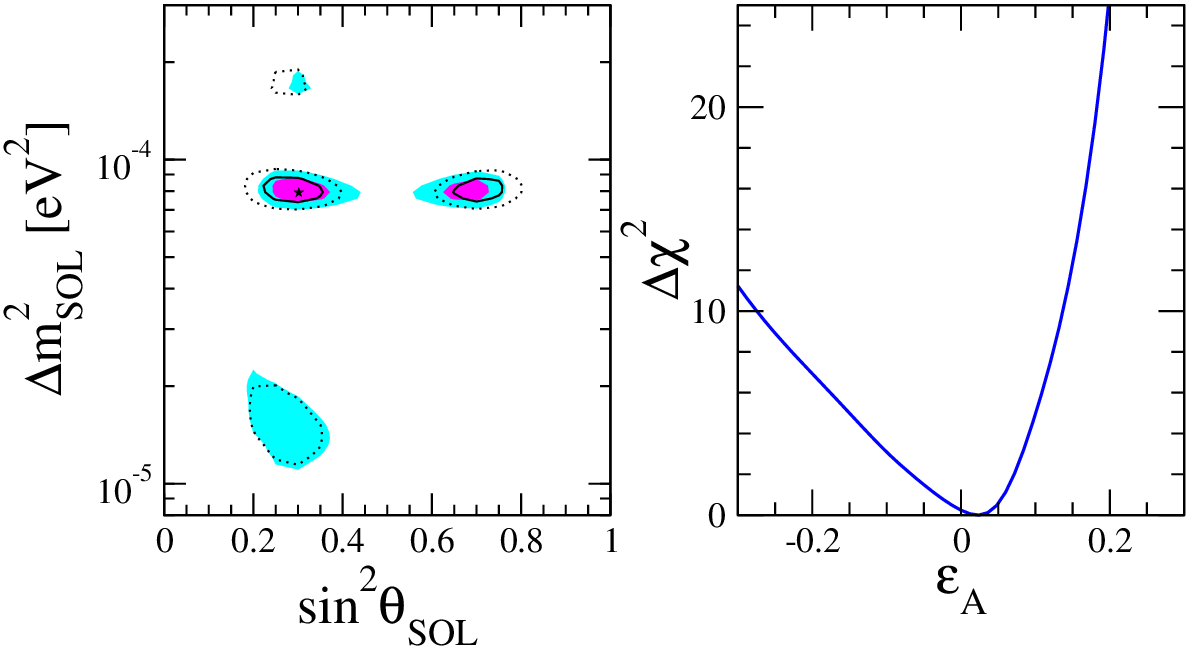}
\caption{\label{fig:eps_a} 
  Results of the analysis including the axial component of the NSI. In
the left panel we compare the allowed regions at 90\% and 3$\sigma$
obtained with (color regions) and without (lines) $\eps_A$. In the
right panel we show the dependence of $\Delta\chi^2$ with respect to
the axial coupling $\eps_A$}.
\end{figure}

The results obtained in a generalized analysis which takes into
account the presence of an non-zero axial component of the NSI (5
parameters instead of 4) are summarized in Fig.  \ref{fig:eps_a}.
One sees that the neutrino data clearly prefers $\eps_A \sim 0$, in
agreement with our previous approximation, in Sec.  \ref{sec:fit}.

\section{Conclusions}
\label{sec:conclusions}

In short, we have reanalysed the status of the LMA oscillation
interpretation of the solar neutrino data in a more general framework
where non-standard neutrino interactions are present.
We have seen that combining the solar neutrino data, including the
latest SNO fluxes of the salt phase with the full KamLAND data sample
still leaves room for a degenerate determination of solar neutrino
oscillation parameters. To this extent the solar neutrino oscillation
parameters extracted from the experiments may be regarded as
non-robust.  In addition to the lower LMA-0 solution, we have found a
LMA-D solution characterized by values of the solar mixing angle
larger than $\pi/4$. This solution requires large non-universal
neutrino interactions on down-type quarks. While the LMA-0 solution is
already disfavored, and will soon be in conflict with further data,
e.g. future KamLAND reactor data, the degeneracy implied by LMA-D
solution will not be resolved by more precise KamLAND reactor
measurements. 
This shows that the determination of solar neutrino parameters only
from solar and KamLAND data is not fully robust.  It is crucial to
consider other data samples, such as atmospheric and laboratory data,
since these bring complementary information. In the present case they
allow one to rule out the LMA-D solution induced by the simplest NSI
between neutrinos and down-type-quarks-only, given the large values of
the non--universal NSI couplings required by that solution.
It is therefore important to perform similar analyses for the more
general case of non-standard interactions involving electrons and/or
up-type quarks.  Only in such scenario (NSI with u-type, d-type and
electrons) we can confidently establish the robustness of the
oscillation interpretation. 
Further experiments, like low-energy solar neutrino experiments
are therefore required in order to clear up the situation.\\

Work supported by Spanish grant FPA2005-01269, by European RTN network
MRTN-CT-2004-503369.  OGM was supported by CONACyT-Mexico and SNI. We
thank Michele Maltoni and Timur Rashba for useful discussions.

%%%%%%%%%%%%%%%%%%%%%%%%%%%%%%%%%%%%%%%%%%%%%%%%%%%%%%%%%%%%%%%%%%%%

\end{document}